\documentclass{iopjournal}

\expandafter\let\csname equation*\endcsname\relax
\expandafter\let\csname endequation*\endcsname\relax

\usepackage{lipsum} 
\usepackage{color}
\usepackage{graphicx}
\usepackage{float}
\usepackage{iopams}
\usepackage{bbding}
\usepackage[margin=10pt,font=small,format=hang]{caption}
\usepackage{amsmath,amssymb}
\usepackage{acronym}
\usepackage{multirow}
\usepackage{tabu}
\usepackage{tikz}
\usetikzlibrary{arrows,shapes,trees,decorations.pathreplacing}
\usepackage{import}
\usepackage{svn-multi}
\usepackage{lineno}
\usepackage{hyperref}
\usepackage{mathtools}
\usepackage{xspace}
\usepackage{tabularx}
\usepackage{url}
\usepackage{cite}
\usepackage{lineno}
\hypersetup{
    colorlinks=true,
    linkcolor=blue,
    filecolor=magenta,      
    urlcolor=cyan,
}

\DeclareGraphicsExtensions{%
    .pdf,.PDF,%
    .png,.PNG,%
    .jpg,.mps,.jpeg,.jbig2,.jb2,.JPG,.JPEG,.JBIG2,.JB2}

\newcommand{\red}[1]{{#1}}

\usepackage{color}

\makeatletter
\newcommand\footnoteref[1]{\protected@xdef\@thefnmark{\ref{#1}}\@footnotemark}
\makeatother

\begin{document}

\articletype{Paper} 

\title{Rapid data quality investigations of gravitational-wave events with the Data Quality Report Builder toolkit}
\author{Derek Davis$^{1,2,*}$\orcid{0000-0001-5620-6751},
        Zach Yarbrough$^{3}$,
        Joseph Areeda$^{4}$,
        Ronaldas Macas$^{5}$, 
        Nicolas Arnaud$^{6}$, 
        Adrian Helmling-Cornell$^{7}$, 
        Paolina Doliva$^{8}$, 
        Olivia Godwin$^{2}$, 
        Hirotaka Yuzurihara$^{9}$, 
        Benjamin Mannix$^{10}$, 
        Sofia \'Alvarez-L\'opez$^{11}$, 
        Max Trevor$^{12}$, 
        Rachael Huxford$^{13}$, 
        Philippe Nguyen$^{10}$,
        Beverly Berger$^{14}$, 
        Chayan Chatterjee$^{15}$, 
        Francesco Di Renzo$^{6}$, 
        Christiano Palomba$^{16}$,
        Viola Sordini$^{6}$, 
        Dimitrios Pesios$^{17}$, 
        Marissa Walker$^{8}$,
        Airene Ahuja$^{18}$,
        Man Leong Chan$^{18}$,
        Julian Ding$^{18}$,
        Raymond Frey$^{10}$,
        Franz Herbst$^{18}$,
        Yannick Lecoeuche$^{18}$,
        Annudesh Liyanage$^{18}$,
        Jess McIver$^{18}$,
        Raymond Ng$^{19}$,
        Sophie Perry $^{18}$,
        Caitlin Rawcliffe$^{18}$,
        Robert Schofield$^{10}$
        }

\affil{$^{1}$University of Rhode Island, Kingston, RI 02881, USA}

\affil{$^{2}$LIGO, California Institute of Technology, Pasadena, CA 91125, USA}

\affil {$^{3}$Louisiana State University, Baton Rouge, LA 70803, USA }

\affil {$^{4}$California State University Fullerton, Fullerton, CA 92831, USA}

\affil{$^{5}$University of Portsmouth, Portsmouth, PO1 3FX, United Kingdom}

\affil {$^{6}$Universit\'e Claude Bernard Lyon 1, CNRS, IP2I Lyon / IN2P3, UMR 5822, F-69622 Villeurbanne, France}

\affil {$^{7}$Bard College, 30 Campus Rd, Annandale-On-Hudson, NY 12504, USA}

\affil {$^{8}$Christopher Newport University, Newport News, VA 23606, USA}

\affil {$^{9}$Institute for Cosmic Ray Research, KAGRA Observatory, The University of Tokyo, 238 Higashi-Mozumi, Kamioka-cho, Hida City, Gifu 506-1205, Japan}

\affil {$^{10}$University of Oregon, Eugene, OR 97403, USA}

\affil {$^{11}$LIGO Laboratory, Massachusetts Institute of Technology, Cambridge, MA 02139, USA}

\affil {$^{12}$University of Maryland, College Park, MD 20742, USA}

\affil {$^{13}$The Pennsylvania State University, University Park, PA 16802, USA}

\affil {$^{14}$Stanford University, Stanford, CA 94305, USA}

\affil {$^{15}$Vanderbilt University, Nashville, TN 37235, USA}

\affil {$^{16}$INFN, Sezione di Roma, I-00185 Roma, Italy}

\affil {$^{17}$Department of Physics, Aristotle University of Thessaloniki, 54124 Thessaloniki, Greece}

\affil{$^{18}$The Department of Physics and Astronomy, The University of British Columbia, Vancouver, BC, V6T 1Z1, Canada}

\affil{$^{19}$The Department of Computer Science, The University of British Columbia, Vancouver, BC, V6T 1Z1, Canada}


\affil{$^*$Author to whom any correspondence should be addressed.}

\email{derek.davis@ligo.org}


\begin{abstract}

We present the Data Quality Report Builder toolkit, \texttt{DQRbuild}, a suite of data quality tools that have been developed to vet gravitational-wave events in preparation for the fourth LIGO-Virgo-KAGRA observing run. 
We explain the main functionality and the many scientific tests that we support.
To validate the performance of the tools included in the toolkit, we run a series of tests on all significant candidates shared as public alerts in the third observing run to compare against what was manually reported using human intervention.
We find that these automated tools can now identify \red{96\%} of the problems identified by humans during this previous observing run, with a \red{24\%} false alarm rate.
We conclude with a commentary on the prospects and potential challenges for fully automating the process of vetting the data quality for gravitational-wave events identified in future observing runs. 

\end{abstract}

\section{Introduction}\label{sec:intro}

Advances in \ac{GW} detector technology~\cite{Capote:2024rmo,VIRGO:2023elp,KAGRA:2022fgc} have pushed the limits of what is possible for the detection of new and exciting \ac{GW} events~\cite{GW150914_paper,GW170817,GWTC-1,GWTC-2,GWTC-2.1,GWTC-3,GWTC-4}. 
One constant challenge of \ac{GW} astronomy is the highly non-stationary and non-Gaussian nature of detector data due to influences of the detector's local environment and non-ideal performance of the detector instrumentation~\cite{Davis:2022dnd,Capote:2024mqe}. 
These \ac{DQ} issues can create false alarms in search pipelines that could be mistaken for an astrophysical signal~\cite{TheLIGOScientific:2017lwt,LIGO_O3_detchar} 
or hamper analysis of real \ac{GW} signals by violating the assumptions of stationarity and Gaussianity used in standard data analysis approaches~\cite{LIGOScientific:2019hgc,Powell:2018csz,Kwok:2021zny,Mozzon:2021wam,Macas:2022afm,Hourihane:2022doe,Ghonge:2023ksb,Davis:2022ird}.
To ensure that astrophysical analyses of \ac{GW} data are both accurate and precise, these \ac{DQ} issues must be actively monitored and accounted for at all stages of the \ac{GW} data analysis process~\cite{LIGO_O3_detchar, Virgo_O3_detchar, KAGRA_detchar}.

In order to address these challenges, it is standard to apply numerous \ac{DQ} checks for any potential candidates, referred to as \emph{event validation}. 
The earliest event validation procedures were applied for the blind injections in pre-detection era to investigate the possibility that these candidates were astrophysical in origin~\cite{detection-checklist-GWDAW07, Abadie:2010yb,Colaboration:2011np}. 
These checks were fundamental for the discovery of GW150914 and the confident claim that this candidate could not possibly be caused by a non-astrophysical source~\cite{GW150914_paper,GW150914_detchar}.
Since this first detection, the approach has changed significantly. 
In the second observing run (O2, from November 2016 to August 2017), these procedures were designed and conducted on an event-by-event basis, as described in GWTC-1.0~\cite{GWTC-1}.
While effective, these early approaches to event validation were highly time-intensive and could therefore be used to validate only a small number of events. 

To address this challenge for the third observing run (O3, from April 2019 to March 2020), the LIGO-Virgo-KAGRA (LVK) collaboration introduced the first tools specifically designed to standardize, manage, and automate these event validation checks~\cite{LIGO_O3_detchar, Virgo_O3_detchar_tools}.
The collated event validation results produced by these tools (the \texttt{data-quality-report}~\cite{DQRdocumentation,LIGO_O3_detchar} and \texttt{VirgoDQR}~\cite{Virgo_O3_detchar_tools} packages) are referred to as \acp{DQR}.
These two tools were used to validate all candidates described in GWTC-2.0, -2.1, and -3.0, both in low latency and as part of the additional astrophysical analyses presented in these catalogs~\cite{GWTC-2,GWTC-2.1,GWTC-3}.
As the detection rate of \acp{GW} has consistently increased with each observing run, these \ac{DQR} tools have become a fundamental part of the \ac{GW} data analysis infrastructure.
Additional details on event validation procedures for detected \ac{GW} signals and how \ac{DQR} results fit into the data analysis ecosystem that alerts astronomers of candidates detected in low latency can be found in~\cite{Davis:2022dnd, Chaudhary:2023vec, DiRenzo:2022klz, Arnaud:2022iiy, DiRenzo:2024xcp}.

Prior to the fourth observing run (O4, from May 2023 to November 2025), we have worked to develop the Data Quality Report Builder (\texttt{DQRbuild}) toolkit to simplify the production of DQRs, building on lessons learned from previous observing runs. 
Key highlights of the code packages that make up this toolkit include new quantitative tests, increased automation, and support for analyses across the international LVK \ac{GW} network. 
In O4, this toolkit was used as a part of the event validation procedure for all transient \ac{GW} candidates identified in low latency using data from the LIGO, Virgo, and KAGRA detectors~\cite{LIGO:2024kkz, GWTC-4_methods}.

In this work, we present an overview of this new toolkit, outline the scientific motivation for its development, and demonstrate the performance of the included science tasks using O3 data.
Finally, we discuss the potential future of this research area as the rate of detection and demand for precision grows.

\section{Data quality goals}

As \ac{DQ} issues can significantly bias analyses of \ac{GW} events or present as false alarms in detection pipelines, it is critical that any potential \ac{DQ} impacts are identified and mitigated before detailed astrophysical analyses are completed, often on the timescale of months to years. 
In particular, when a \ac{GW} event is identified in low latency, there is a need for rapid investigations into whether the information provided about the candidate may be biased by \ac{DQ} issues, or if the candidate is actually not astrophysical in origin.
To maximize the science that can be done with gravitational-wave candidates, we must balance the need for thorough, accurate analyses with the need for rapid communication and decision-making.
To motivate the design choices in the \texttt{DQRbuild} toolkit, we will examine the \ac{DQ} needs and latency requirements for both astrophysical and non-astrophysical candidates. 

\subsection{Astrophysical candidates with DQ issues}

For real signals, the impact of any \ac{DQ} issues that may be present near the signal is highly dependent on the rate and type of \ac{DQ} issue present in the specific data. 
What can be said is that the rate of these \ac{DQ} issues is non-trivial, impacting more than 10\% of all \ac{GW} events announced in GWTC-3.0 and GWTC-4.0 with false alarm rates less than 1/yr~\cite{GWTC-3, GWTC-4}. 

As an ad-hoc estimate of how often we expect these types of \ac{DQ} issues to be present, we can calculate the rate of coincidence between \ac{GW} signals and glitches based on the length of data used when analyzing an event and the rate of glitches. 
In O3, a typical rate of glitches with SNR $> 6.5$ was 1 per minute~\cite{GWTC-3}. 
While recent work has shown that glitches must directly overlap the \ac{GW} signal in question to cause biases~\cite{Hourihane:2025vxc}, the standard approach in analyses has been to mitigate any glitches present during the analysis period.
Assuming that an analysis of a typical binary black hole signal uses 8\,s of data in parameter estimation analyses, and that the number of glitches in a given interval follows a Poisson distribution, there is a \red{12\%} chance of a glitch occurring in a single detector during the 8\,s overlapping the analysis window. 
As multiple detectors are used to analyze signals, the actual rate of glitch overlaps will be higher. 
For example, if three detectors are observing coincidentally, there is a \red{32\%} chance that at least one of them contains a glitch during this time span.
This value is consistent with the rate of \ac{DQ} issues in O3 events with parameter estimation results reported in GWTC-2.0 and GWTC-3.0, where \red{18} of \red{74} events (\red{25\%}) experienced \ac{DQ} issues~\cite{GWTC-2,GWTC-3}. 

Most \ac{DQ} issues that are flagged in this category are straightforward to identify.
A wide variety of tests are available to determine if data is consistent with Gaussian noise plus an astrophysical signal, from tests of anomalous power in spectrograms~\cite{Vazsonyi:2022jul,Mozzon:2020gwa} and machine-learning classifiers of spectrogram data~\cite{Zevin:2016qwy,Alvarez-Lopez:2023dmv} to statistical hypothesis tests~\cite{Cornish:2014kda, Yamamura:2024vka}.
This process can also be performed through the inspection of data by hand~\cite{ref:omegagrams}; however, these types of visual checks are highly subjective, and significant differences have been found between experts in how they identify non-Gaussian power, as well as between visual inspection and the aforementioned statistical tests~\cite{Vazsonyi:2022jul}.
Due to the wide variety of tools available, the primary challenge is tuning and automating these tests to address the diverse range of both potential signal morphologies and potential \ac{DQ} issues.

In addition to the clear \ac{DQ} issues that can be identified using Gaussianity tests, recent work has shown that more subtle differences between real \ac{GW} data and Gaussian noise can create biases in the analysis of \ac{GW} events~\cite{Payne:2022,Macas:2023wiw,Udall:2025bts,Ray:2025rtt}.
Tools that utilize auxiliary channel information to calculate if non-stationary noise is present in the data are the primary method available now to identify these types of subtle noise sources~\cite{AdvLIGO:2021oxw, Helmling-Cornell:2023wqe}. 
These tools rely upon significant, ongoing efforts at the observatory sites to acquire the relevant knowledge to accurately compute these noise contributions, which limits the total number of tools available in this category.
Work is ongoing to develop additional tools to identify and mitigate these types of subtle \ac{DQ} issues. 

\subsection{Non-astrophysical candidates}

The other scenario we are concerned about is if a candidate identified by a search pipeline is not of astrophysical origin. 
Generally, this can occur for two reasons: a non-zero rate of false alarms is expected by definition of the false alarm rate assigned to each candidate, or incorrect assumptions are made when calculating that false alarm rate. 
The latter case is particularly relevant for low-latency analyses, as the analysis of real-time data requires searches to assume that past data is representative of future data, despite the high level of non-stationary noise behavior that \ac{GW} detectors are known to exhibit~\cite{Nitz:2018rgo, Messick:2016aqy, Adams:2015ulm, Chu:2020pjv}. 
The high-risk, high-reward nature of low-latency searches makes it crucial that information about any non-astrophysical candidates is provided promptly. 
Identifying non-astrophysical signals for high-latency searches is also essential, as it has been observed that non-astrophysical candidates are much more likely to be consistent with non-standard \ac{GW} parameters~\cite{Heinzel:2023vkq,Ruiz-Rocha:2025yno}; if the non-astrophysical nature of these candidates is not identified, these potentially exceptional cases could significantly pollute our understanding of \ac{GW} sources. 

Although we must identify any non-astrophysical candidates, we have a high standard of evidence for declaring something not to be astrophysical in origin (particularly in low latency). 
This is motivated by the desire not to incorrectly discard a real signal. 
At the same time, we aim to thoroughly investigate if a signal could be non-astrophysical to increase our confidence in the detected signals. 
This process was one of the key components of declaring GW150914 the first direct detection of \acp{GW}~\cite{GW150914_detchar}.
The exclusion of other non-astrophysical explanations strengthened the case for the astrophysical origin of this signal.

There are two primary avenues to build a case for a candidate being non-astrophysical in origin: either the measured strain data is incompatible with the claimed candidate, or the signal in the strain data related to the claim can be explained by a known instrumental coupling that is supported by additional auxiliary channel information.
The former case can be established through the investigation of the strain data alone, using residual analyses and machine-learning classifiers mentioned in the previous section. 
For the latter scenario, we rely on the thousands of additional auxiliary sensors located at each \ac{GW} observatory to monitor the detector performance and its environment~\cite{AdvLIGO:2021oxw,Virgo:2022ypn}.
Correlation between an auxiliary sensor and the strain data can be based on statistical analyses of both data streams~\cite{Smith:2011an,Essick:2020qpo} or through physical injections at the observatory sites to measure the coupling between different auxiliary sensors and strain data~\cite{AdvLIGO:2021oxw, Helmling-Cornell:2023wqe}.

\subsection{Latency requirements}

Although it would be desirable for all investigations to be completed as rapidly as possible, limitations on computing resource limitations as well as on data availability mean that certain types of \ac{DQ} investigations must be prioritized over others. 
To address these issues, we consider three successive latency tiers with different science goals in mind:

\begin{itemize}
    \item{\textbf{5 minutes:} This is the latency required for \ac{DQ} information to be useful for the majority of retractions that occur in low latency.
     Based on the O3 experience, the minimum time for manually retracting or confirming a single event was approximately 20 minutes~\cite{G2001971}.
     A latency of 5 minutes allows for \ac{DQ} results to be used as part of a retraction or confirmation process that occurs within 10 minutes of when the events was first detected.
     A latency of 5 minutes is also the lowest that can be reasonably and reliably achieved by current analyses in the \texttt{DQRbuild} toolkit, 
     as latencies that are part of the infrastructure, such as the time to be notified of an event and start analyses,
     generally takes multiple minutes~\cite{Sachdev:2020lfd,Magee:2021xdx,Chaudhary:2023vec}. }
     
    \item{\textbf{1 hour:} This latency is required for input on a small number of low-latency retractions that require additional time to evaluate.
     In O3, a small number of candidates were retracted on this timescale~\cite{GraceDB}. 
     A latency of 1 hour also corresponds to the approximate amount of time that many wide-field survey telescopes require to initiate target-of-opportunity observations
     of \ac{GW} candidates and the expected rising timescale of thermal emission from kilonovae~\cite{Coughlin:2019xfb, DECam:2020qlw, Ahumada:2024qpr, Andreoni:2021epw}. }
     
    \item{\textbf{1 day:} This latency is required for analysis candidates that may require additional vetting, but are still of interest for time-sensitive follow up by other observatories.
     In O3, a single trigger fell into this category of candidates that were detected in low latency~\cite{GraceDB}. 
     The entire sky-localization area of many \ac{GW} candidates can be observed with electromagnetic telescopes within 1-2 days~\cite{Coughlin:2020fwx}, 
     while additional follow up observations can continue for multiple days~\cite{DES:2020ref}. 
     Furthermore, the kilonova associated with GW170817 peaked in brightness after 1-2 days 
     and decayed in magnitude at a rate of 1 mag per day ~\cite{Drout:2017ijr,Kasliwal:2020wmy}.
     This maximum magnitude and rate of decay means that most observable candidates will 
     only be able to be identified for a few days. 
     Therefore, 1 day is in practice the longest latency for event validation analyses that optimizes electromagnetic follow-up 
     and minimizes delays to other \ac{GW} analyses.}
\end{itemize}

We do note that many telescopes have latencies of less than 5 minutes (see, e.g.,~\cite{CHIMEFRB:2018mlh,James:2019xca,Tohuvavohu:2024flg}) or respond to \ac{GW} candidates that are identified before merger~\cite{Magee:2021xdx}.
However, the low latency of these telescopes that allows such observations to be possible also means that the overall time spent on each target of opportunity observation is lower, minimizing the impact of follow up of any non-astrophysical candidates. 
Furthermore, many of these telescopes with low latencies may have minimal ability to target specific points in the sky, meaning that the choice to follow up of candidates is only dependent on the telescope coverage at the time of the candidate.
Therefore, the vast majority of science benefits from \ac{DQ} investigation of candidates identified in low latency can be realized by targeting latencies between 5 minutes and 1 day.

\section{Core elements of the DQRbuild toolkit}

In order to create a \ac{DQR} using the \texttt{DQRbuild} toolkit, we need to operate a wide variety of different technical tasks at various sites across the global \ac{GW} network. 
The amount of data consumed by these analyses is much greater than is typically transmitted across the global network in low latency, making it impractical at this time to transport the input data used by the \texttt{DQRbuild} toolkit to a single location. 
Hence, the \texttt{DQRbuild} toolkit must be able to operate in a wide variety of different environments and be able to adapt to the needs of the local analyses. 
Each location where the \texttt{DQRbuild} toolkit is implemented is referred to as a ``\texttt{DQRbuild} node.''

\begin{figure}[t]
\centering
\includegraphics[width=0.85\textwidth]{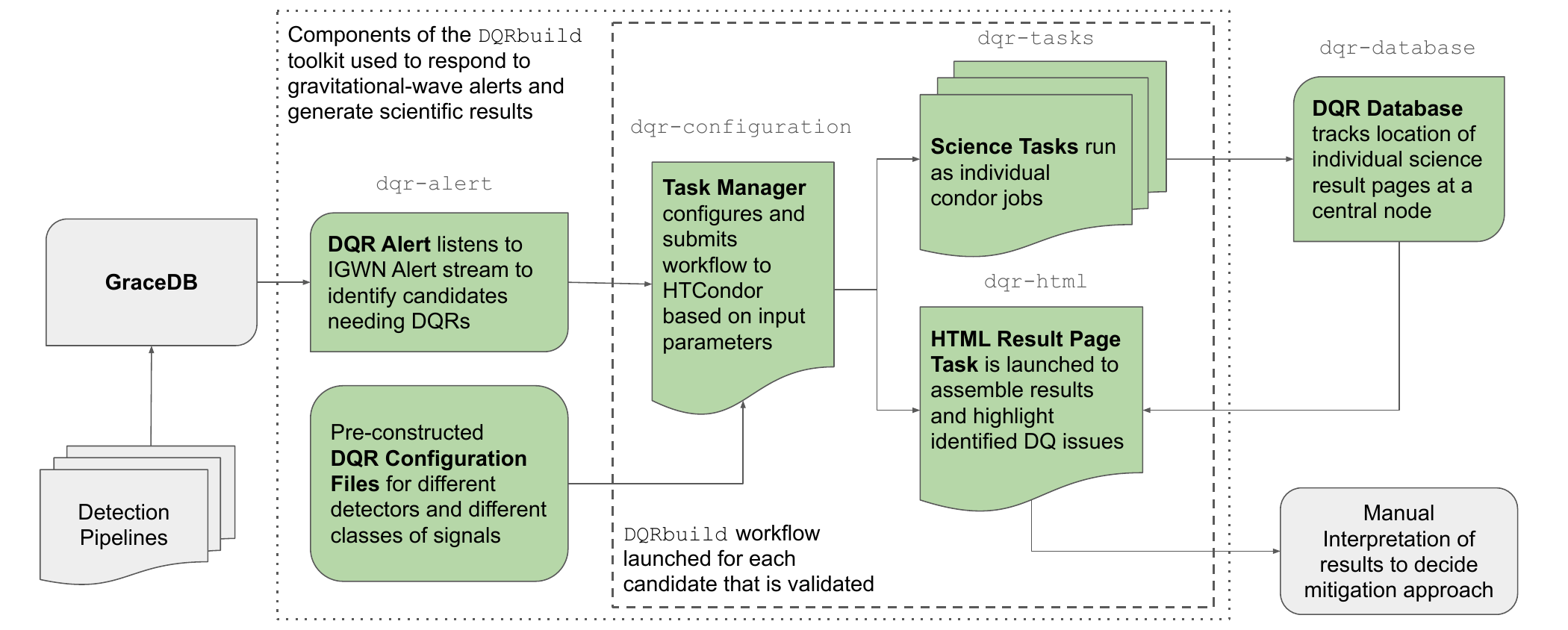}
\caption{A visualization of the inter-connected components of the \texttt{DQRbuild} toolkit implemented at a single \texttt{DQRbuild} node, along with other related databases.
Shown is the process in which the \texttt{dqr-alert} listener identifies an alert from the Gravitational-Wave Candidate Event Database (GraceDB)~\cite{GraceDB} and then initiates \texttt{dqr-configuration} to launch a new \texttt{DQRbuild} workflow. 
These analyses are then completed using HTCondor and tracked with \texttt{dqr-database}.}
\label{fig:manager}
\end{figure}

The infrastructure of the \texttt{DQRbuild} toolkit is split into several sub-packages with specific goals, namely \texttt{dqr-configuration}, \texttt{dqr-alert}, \texttt{dqr-database}, and \texttt{dqr-tasks}. \
The full workflow between each of these sub-packages (as well as interactions with other outside resources) is shown in Figure~\ref{fig:manager}.
We refer to the analysis that is launched using the \texttt{DQRbuild} toolkit to create a \ac{DQR} as a ``\texttt{DQRbuild} workflow.''

\subsection{dqr-configuration}

This is the ``task manager'' that is used to create the desired \texttt{DQRbuild} workflow. 
As the \texttt{DQRbuild} toolkit relies on HTCondor~\cite{condor-hunter,Couvares2007}, designing the workflow is synonymous with creating the relevant files that describe how each element of the workflow will be run along with a directed acyclic graph (DAG) file that describes the parent-child relationships between each element.
We refer to the individual elements in a \texttt{DQRbuild} workflow as ``tasks.'' 

The \texttt{dqr-configuration} package relies on configuration files that describe the contents of each portion of the workflow. 
In this way, there is extreme flexibility in how this toolkit can be run, as the specific implementation is coordinated via these user-created configuration files rather than hard-coded components of the package. 

This script also includes features such as command-line configuration overrides to allow for easy changes to a candidate workflow and setting variables in the workflow that are based on either provided event JSON files or information downloaded from the \ac{GW} Candidate Database (GraceDB).

\subsection{dqr-alert}

The \texttt{dqr-alert} package is designed to take advantage of the \texttt{igwn-alert} infrastructure to launch independent programs, such as the \texttt{dqr-configuration} task manager, when an appropriate alert is issued by GraceDB. 
The \texttt{igwn-alert} package is, in turn, a client for the LIGO-Virgo-KAGRA alert infrastructure that uses the Apache Kafka framework to alert users and follow-up processes to state changes in GraceDB. Details about the \texttt{igwn-alert} tool can be found in \cite{igwn-alert-documentation}.

The \texttt{dqr-alert} structure is particularly important for coordinating analyses using infrastructure in the \texttt{DQRbuild} toolkit that is spread out across multiple detector sites. 
This set up allows automated actions to be taken synchronously across the worldwide detector network.

\subsection{dqr-database}

The \texttt{dqr-database} package is designed to assist with tracking the large number of results that are uploaded from individual detector sites to a centralized results location.
That locally-hosted database relies on SQL.
The database tracks the versions of each local \texttt{DQRbuild} instance and the location of all individual task results (generated locally or uploaded from external computing centers) to facilitate the generation of result pages that summarize the complete set of event validation analyses for each event. 
This information is used by the tools that create HTML pages collating the scientific task results to correctly link the central page to all of the relevant results.

\subsection{dqr-tasks}

The \texttt{dqr-tasks} package implements numerous tasks intended to be used as part of a \texttt{DQRbuild} workflow. 
This includes ``science tasks'' that investigate a specific \ac{DQ} issue around an event of interest and ``utility tasks'' that complete other critical operations. 
Science tasks will be described in detail in Section~\ref{sec:science}.
Examples of utility tasks include scripts that transfer results between different sites, create HTML pages to display the results, find required data, or convert output file formats.
The first two of these examples are of particular importance for the use of the \texttt{DQRbuild} toolkit in event validation analyses, and are explored more in this section. 

Tasks used in a \texttt{DQRbuild} workflow are not required to be implemented in this package. 
Instead, there is a standard JSON format that tasks use to store results. 
Other packages that produce results in this format (or rely on scripts to convert their results into this format) can be understood by other elements of a \texttt{DQRbuild} workflow. 

\subsubsection{dqr-html}

\begin{figure}[t]
\centering
\includegraphics[width=0.95\textwidth]{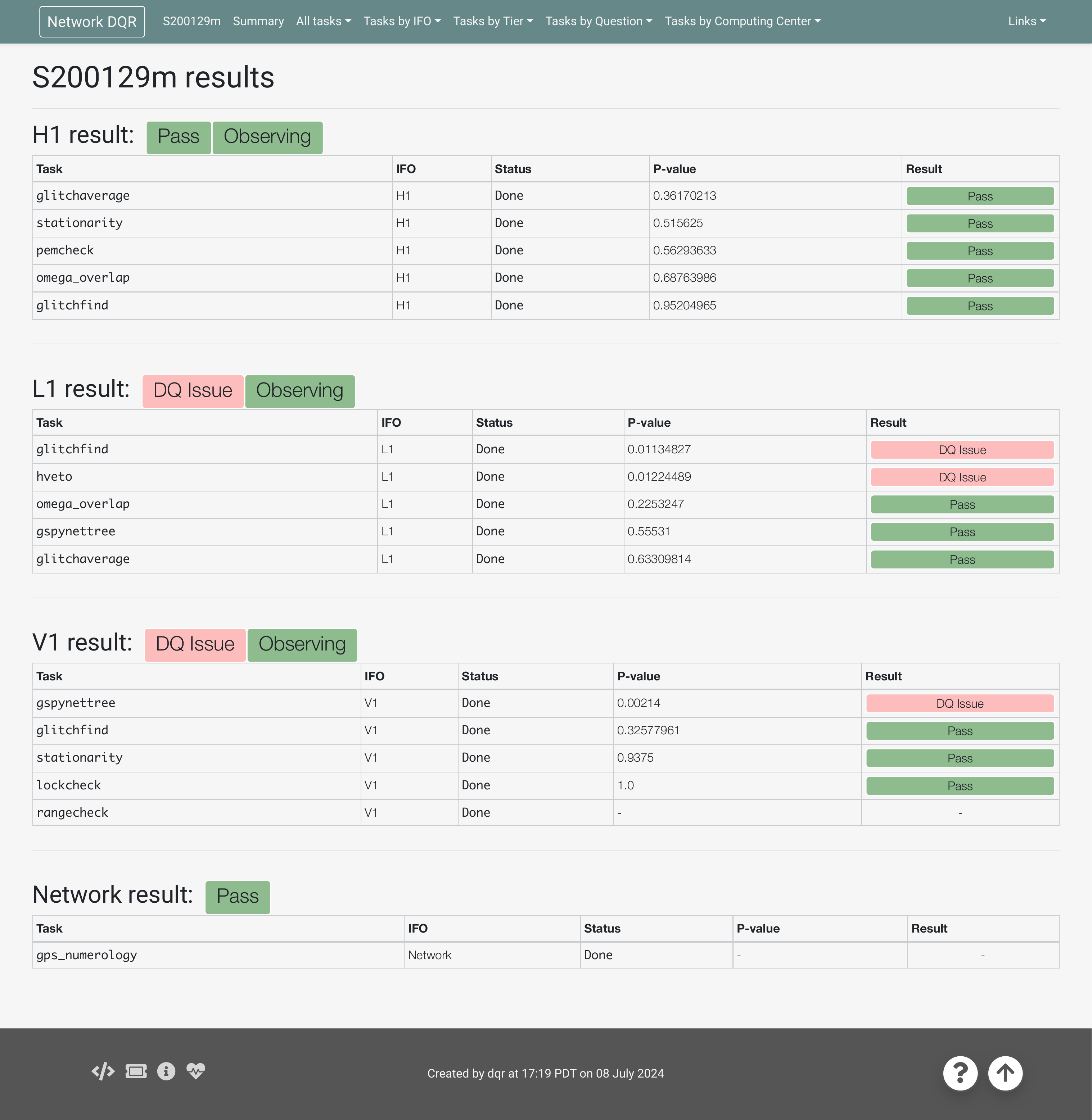}
\caption{An example HTML results page for the candidate S200129m.
This example highlights the ease with which relevant \ac{DQ} information is accessed from the \texttt{DQRbuild} result pages.}
\label{fig:html}
\end{figure}

This script (implemented in the \texttt{dqr-tasks} package as \texttt{dqr-html}) creates the HTML pages that are used to display the results and to identify which tasks returned \ac{DQ} issues. 
These pages include a front page, which shows a small number of tasks from each detector, as well as pages that display tasks from a specific interferometer or latency tier.
An example of the result page produced by this task is shown in Figure~\ref{fig:html}.

This example shows the results for \red{S200129m}.
In this single image, a user can see that all three detectors are observing and that some tasks identified \ac{DQ} issues in the LIGO Livingston and Virgo interferometers, consistent with what was previously reported from manual vetting of this data~\cite{GWTC-3, Payne:2022}.
The displayed table also includes links to the result page and HTCondor log messages for each individual task.
This allows quick interpretation of the result and makes it straightforward to investigate any computing errors from a web browser. 
Further details about each task, as well as visualizations of all tasks sorted into different categories can be accessed by the pull-down menus at the top of the page.

\subsubsection{dqr-upload}

This script (implemented in the \texttt{dqr-tasks} package as \texttt{dqr-upload}) that compresses the result files and then transfers them between different nodes using \texttt{scp} protocols. 
After each upload is completed, the script then runs an additional command to update the \texttt{dqr-database} at the central node to ensure that the result is properly tracked. 
While not necessary for each individual \texttt{DQRbuild} node to function, this tool is essential for the federated approach taken with respect to the use of \texttt{DQRbuild} across multiple computing sites.

\subsection{Use of DQRbuild toolkit elements during the O4 run}

As described in GWTC-5~\cite{GWTC-5_methods}, the \texttt{DQRbuild} toolkit was extensively used to validate candidates identified in O4.
This included a central \ac{DQR} node that hosted a \texttt{dqr-database} instance, as well as additional nodes at the LIGO Hanford, LIGO Livingston, and Virgo observatories. 
All nodes relied upon \texttt{dqr-alert} to autonomously launch \ac{DQR} workflows, and all nodes but the central node used \texttt{dqr-upload} to transfer results to the central node and log them in the relevant database (there was no need to upload from the central node). 

The structure of the \ac{DQR} workflow that was launched at each node was heterogeneous across the network.
The central node, the LIGO Livingston node, and the LIGO Hanford node were used to launch \texttt{DQRbuild} workflows using \texttt{dqr-configuration}. 
These workflows included tasks from the \texttt{dqr-tasks} and \texttt{GWDetchar}~\cite{gwdetchar} packages. 
The node at Virgo relied upon \texttt{VirgoDQR} to launch workflows that included tasks from the \texttt{dqr-tasks} and \texttt{VirgoDQR} packages.

The manual event-validation process that ends the vetting phase of all candidates prior to using them in offline analyses was also initiated by a task in the \texttt{event-validation} package~\cite{LIGO:2024kkz} at the central node.

Note that data from each interferometer in the LVK network were analyzed by multiple \ac{DQR} nodes.
The central node analyzed data from all four interferometers in the LVK network; the Virgo node analyzed data from all interferometers except KAGRA; and the LIGO Hanford and LIGO Livingston nodes analyzed data from their local interferometers. 

\section{Science tasks} \label{sec:science}

To address the need for automation and the desire to include as much information as possible in the \texttt{DQRbuild} toolkit, we categorized scientific tasks into three categories: statistical tasks, binary tasks, and qualitative tasks. 
We now look at the tasks that fit into each of these categories that have been implemented into the \texttt{dqr-tasks} package.

\subsection{Statistical tasks}

Tasks that report a \textbf{p-value} are considered ``statistical tasks.''
The primary goal of these tasks is to report a p-value that indicates the probability that the task result is incorrectly identified as a \ac{DQ} issue relevant to the candidate. 
Tasks producing results in this \textbf{p-value} format will allow automated logic boxes to iterate over the tasks and determine whether:
\begin{itemize}
    \item{the candidate passes,
        \begin{itemize}
            \item{all tasks report a \textbf{p-value} above a chosen, task-dependent threshold, 
                  i.e., zero tasks identified a statistically-significant \ac{DQ} concern}
        \end{itemize}
    }
    \item{the candidate has a \ac{DQ} issue,
        \begin{itemize}
            \item{at least one task reports a \textbf{p-value} below a chosen, task-dependent threshold, 
                  i.e., at least one task identified a statistically-significant \ac{DQ} concern}
        \end{itemize}
    }
\end{itemize}
Note that some results are truly binary, e.g., whether the detector was observing or not. 
These tasks are considered ``binary tasks.''
In such a binary case, it would be possible to consider each binary result as a \textbf{p-value} of 1 or 0. 

\subsubsection{HVeto} \label{sec:hveto}
This task (implemented in the \texttt{dqr-tasks} package as \texttt{dqr-hveto}) is designed to identify whether a candidate \ac{GW} event coincides with statistically significant transient noise in auxiliary or environmental sensors. This task relies on the hierarchical veto algorithm (HVeto) to pinpoint channels that show excess coincidences with glitch-like triggers in the main strain channel~\cite{Smith:2011an}.

The task begins with identifying time intervals of excess noise in the detector's strain channel using an event trigger generator, such as Omicron~\cite{Robinet:2020lbf} or SNAX~\cite{snax-documentation}. The task searches for statistically significant coincident triggers  between the strain channel and auxiliary channels by testing multiple different coincidence time windows and SNR thresholds. Those channels showing significant coincident rates are selected for follow-up.

The task proceeds iteratively: in each round, it chooses the auxiliary channel that contains the most glitches that are coincident with gitches in the strain channel and uses that channel to form veto segments around these times. The task then repeats this process using only un-vetoed strain triggers. This process continues until no auxiliary channel exhibits significant correlation above a configurable significance threshold.

\begin{figure}[t]
\centering
\includegraphics[width=0.95\textwidth]{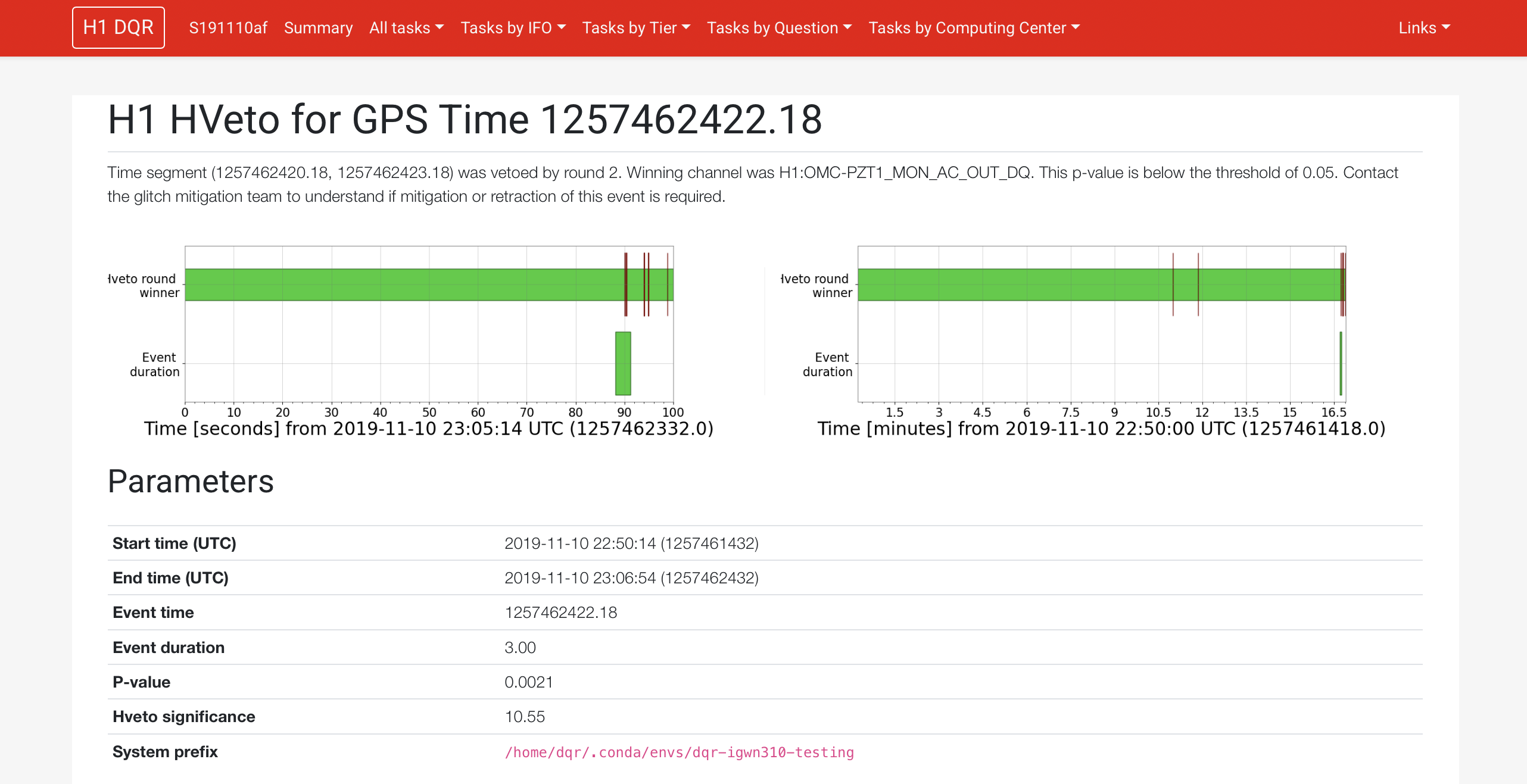}
\caption{An example result page for the HVeto task.
The displayed candidate is S191011af, which was retracted due to a correlation between the observed strain signal and excess noise from a piezoelectric controller (PZT) in the output mode cleaner (OMC) of LIGO Hanford~\cite{LIGO_O3_detchar}.}
\label{fig:hveto}
\end{figure}

An example of this task run on \red{S191110af} is shown in Figure~\ref{fig:hveto}.
In this example, HVeto is able to flag a correlation between the observed strain signal and excess noise from a piezoelectric controller (PZT) in the output mode cleaner (OMC) of LIGO Hanford~\cite{LIGO_O3_detchar}.
This correlation was first identified near the time of the candidate after multiple days of investigations and was used as the evidence needed to retract this \ac{GW} candidate.
The current HVeto task was able to identify this correlation within 5 minutes of when the required input trigger data was available. 

\subsubsection{Stationarity} \label{sec:stationarity}
This task (implemented in the \texttt{dqr-tasks} package as \texttt{dqr-stationarity}) is designed to determine if the noise spectrum during a short-duration signal is elevated compared to other nearby time periods. It has been noted in previous publications that an incorrect measurement of the power spectral density (PSD) can result in a spurious \ac{GW} candidate~\cite{Zackay:2019kkv,Mozzon:2020gwa}. This task is based on the stationarity metric used in the PyCBC~\cite{Usman:2015kfa} search for \acp{GW} from compact binary coalescences~\cite{Mozzon:2020gwa}.

The statistic reported by this task, $\rho$, is calculated with a short-duration PSD (often 8 seconds long), $S_{SD}(f)$, and a long-duration PSD (often 512 seconds long), $S_{LD}(f)$. The statistic is also weighted by the frequency spectrum of a binary inspiral, $f^{-7/3}$. The full equation is as follows:
\begin{equation}
    \rho = \sum_f \frac{S_{LD}(f)}{f^{-7/3}}\times \sum_f \frac{f^{-7/3}S_{SD}(f)}{S_{LD}^2(f)}
\end{equation}
In short, this statistic is a frequency-weighted ratio of the short PSD to the long PSD. The statistic is normalized so that the mean value is 1.

To calculate the p-value reported by this task, data consisting of only colored Gaussian noise is generated with \texttt{bilby}~\cite{Romero-Shaw:bilby:2020} that has the same noise spectrum as the long-duration PSD. The same statistic used in this task is calculated over multiple, non-overlapping time periods in the simulated data with the same duration as the real data. The distribution of values from this test is then compared against the statistic measured using real data. Therefore, this p-value is “the probability that Gaussian noise could produce a stationarity statistic higher than the value measured with real data.”


\subsubsection{iDQ} \label{sec:idq}
This task (implemented in the \texttt{dqr-tasks} package as \texttt{dqr-idq}) is designed to evaluate whether the data surrounding a candidate \ac{GW} event exhibit statistical behavior consistent with instances of transient noise. This task utilizes the information from the inferential Data Quality (iDQ) algorithm that is running online at the LIGO sites~\cite{Essick:2020qpo}.

The core output of this task is the iDQ probability, which estimates the likelihood that the data segment contains a glitch, based on auxiliary channel correlations and past examples of glitches. 
To contextualize this probability, the task compares the value measured at the time of the candidate event to a background distribution computed over a longer surrounding segment. This allows the task to determine whether the event time is unusually glitch-like, given the detector's recent behavior.

The p-value reported by this task is defined as the fraction of times within the background window that have an iDQ probability equal to or greater than that measured at the candidate time. A low p-value therefore indicates that the iDQ classifier identifies a candidate time as more glitch-like than most of its surroundings. 


\subsubsection{PEMcheck} \label{sec:pemcheck}
The PEMcheck task (implemented in the \texttt{dqr-tasks} package as \texttt{dqr-pemcheck}) projects the level of environment noise present in the strain data at the time of an event~\cite{Helmling-Cornell:2023wqe}.
To do so, this task utilizes \emph{coupling functions} between the \ac{GW} interferometers and a network of auxiliary environmental sensors (referred to as the ``physical environmental monitoring'' system, or the PEM system) measured through a systematic noise injection campaign~\cite{AdvLIGO:2021oxw} and couplings calculated by this task from data surrounding the time of interest.

Using these known coupling functions~\cite{pem.ligo.org}, the amplitude of environmental noise in the interferometer strain data can be estimated based on the ambient noise levels in each environmental sensor at the time of the \ac{GW} candidate.
This calculation is typically done on a sensor-by-sensor basis, but for electronics bay magnetometers, we average the relevant statistic over the number of magnetometers as these sensors are expected to be highly correlated. 
Averaging over multiple sensors acts as a trials factor for what amounts to performing a handful of nearly-identical tests for the likelihood of environmental signals contaminating the interferometer strain data. 
The coupling functions used for each candidate are also adjusted based on surrounding ambient noise data; 
in cases where predicted ambient environmental amplitude near (but not overlapping) the time of a candidate exceeds the actual strain amplitude observed, we reduce the coupling function at that point such that the predicted noise amplitude equals the strain amplitude. 

For each environmental sensor channel and the strain channel, we compute the spectrogram near the relevant \ac{GW} candidate.
We also compute the spectrogram of the expected noise (the ``projected spectrogram'') in the strain channel due to noise by first filtering the environmental data based on the adjusted coupling function.
For each time-frequency tile in the projected spectrogram with projected strain $\mu$, we identify the strain $h$ in the corresponding time-frequency tile of the strain spectrogram. We then compute the probability of observing $h$ given $\mu$. 
We call this the $c$-statistic, or contamination statistic. It is designed such that the lower the value, the higher the probability that there is not excess noise in the \ac{GW} strain from any disturbances witnessed by the PEM sensors.
For event validation, we discard all $c$-statistics which were calculated in tiles that are far away from the predicted inspiral track or time-frequency window in the case of a burst-only detection. 
For each channel, we report the minimum $c$-statistic as well as the time and frequency of that tile. 
Additional work is required to convert this $c$-statistic into a properly calibrated p-value. 

In addition to being important for event validation, this task can also be used as a monitor of the stability of PEM coupling functions with respect to time. 
If a specific environmental sensor is flagged by this task as exceeding the background noise level but there is no noise present in the strain data consistent with what the task predicts (based on other tasks), this may be due to changes in the coupling function. 
If such a scenario occurs, this could prompt new measurements of the coupling functions along with investigations of detector logbooks to assess if there is a potential change in the detector response to environmental noise. 


\subsubsection{Omega Overlap} \label{sec:omega_overlap}

The Omega Overlap task (implemented in the \texttt{dqr-tasks} package as \texttt{dqr-omega-overlap}) is designed to test if a given auxiliary channel has excess power coinciding with the candidate event.
If there is coincident power, the false alarm rate of the coincidence is based on the probability of this amount of power being coincident by chance based on the behavior of channel in the surrounding data.
If the data from the auxiliary channel is usually quiet, this could indicate that the auxiliary channel recorded the noise which maybe caused the excess power in h(t).

To identify excess power and estimate its significance, the task performs the following steps:
\begin{itemize}
    \item Find the time-frequency parameters for a given candidate event.
    \item Select 1\% of the most energetic pixels in the time-frequency spectrogram of the auxiliary channel.
    \item Estimate how many of those 1\% pixels are within the time-frequency track of the given candidate event.
    \item Estimate what is the total energy of these 1\% pixels within the time-frequency track of the given candidate event.
    \item Measure how often this auxiliary channel has this amount of energy at this time-frequency volume.
\end{itemize}

As there are typically hundreds to thousands of auxiliary channels monitored to identify coincident witness to each candidate \ac{GW}, care must be taken to estimate the false alarm rate over the full set of channels considered to counteract the multiple comparison problem. 
To address this, this task provides a scaled p-value which takes into account the number of channels that are analyzed. 
The scaled p-value is $p_\text{scaled} = 1 - (1-p)^N$, where $p$ is the lowest p-value reported for any channel analyzed, and $N$ is the number of channels analyzed
(This is also known as the \u Sid\'ak correction~\cite{Sidak01061967}).
In this way, the sensitivity of the task to excess power in any given channel is dependent on the total number of channels that are analyzed. 

\subsubsection{Glitchfind} \label{sec:glitchfind}

This task (implemented in the \texttt{dqr-tasks} package as \texttt{dqr-glitch-find}) is designed to determine whether known noise transients occurred near the time of a \ac{GW} candidate using known statistical properties of Gaussian noise under the Q-transform~\cite{Vazsonyi:2022jul,Chatterji:2004qg}. 

The task identifies statistically significant bursts of excess power by first generating a ``Q-gram'' using \texttt{gwpy}~\cite{2021SoftX..1300657M}, which uses sine-Gaussian wavelets with a fixed Q to transform the time series into the time-frequency space. 
The task then creates histograms of the individual wavelet SNR values and fits the low-SNR bulk data to an exponential. 
This fit is then used to estimate the probability of observing the highest SNR wavelet in this data under the Gaussian noise null hypothesis; this probability estimate is the p-value reported by the task.
As the presence of a \ac{GW} event is likely to be inconsistent with Gaussian noise on its own, the task first subtracts the best-fit template identified by a matched filter search algorithm to ensure that any excess power identified is not from the signal identified by this search. 
The assumption that the best-fit template accurately reflects the real data does introduce a potential source of bias; 
however, this has been identified as a concern in only a few cases of real \ac{GW} signals observed to date.

As it has been shown that glitches not overlapping the time-frequency space of the signal do not bias parameter estimation~\cite{Hourihane:2025vxc}, this task can also be configured to only look for excess power close to the signal track. 
This ensures that glitches far from the signal are not flagged and increases the sensitivity of the task, as there is less time-frequency space to analyze. 
This increased sensitivity makes it more likely that low-SNR bursts, inconsistent with Gaussian noise, are flagged by this task.
However, it is not clear if these types of low-SNR bursts are loud enough to bias analyses of \ac{GW}s.


\subsubsection{GSpyNetTree} \label{sec:gspynettree}

This task (implemented in the \texttt{dqr-tasks} package as \texttt{dqr-gspynettree}) is based upon GSpyNetTree, the Gravity Spy Convolutional Neural Network Decision Tree, which uses machine learning to determine whether a glitch is present at the time of a candidate event~\cite{Alvarez-Lopez:2023dmv}. 
GSpyNetTree leverages a decision tree of multi-label CNN classifiers, sorted via total estimated \ac{GW} candidate mass, and trained with morphologically similar glitches.

The use of a multi-label architecture for the convolutional neural networks in GSpyNetTree means this task also considers cases where a \ac{GW} candidate and a glitch overlap in time (and frequency).
With a multi-label architecture, GSpyNetTree is able to predict 0 or more labels for each candidate, by returning a probability ranging from 0 to 1 for each considered class. 
This way, the sum of the probabilities of all labels is not 1 (as occurs for multi-class classifiers, where the classes are mutually exclusive). 
Instead, the probability of each label can take any value from 0 to 1.
If GSpyNetTree predicts that a glitch is present (including the case where a \ac{GW} and/or No-Glitch label is simultaneously predicted with a glitch), GSpyNetTree needs to determine if a \ac{DQ} issue should be flagged. 
To do this, GSpyNetTree uses the p-value of the data not containing a glitch from any class, which ranges from 0 (\ac{DQ} issue identified) to 1 (no \ac{DQ} issue identified). 
This p-value is calculated as one minus the maximum probability that the data is consistent with a specific glitch class across all classes considered, such that if the probability of the glitch is very high, the p-value will be near zero and a \ac{DQ} issue will be flagged. 

\begin{figure}[t]
\centering
\includegraphics[width=0.95\textwidth]{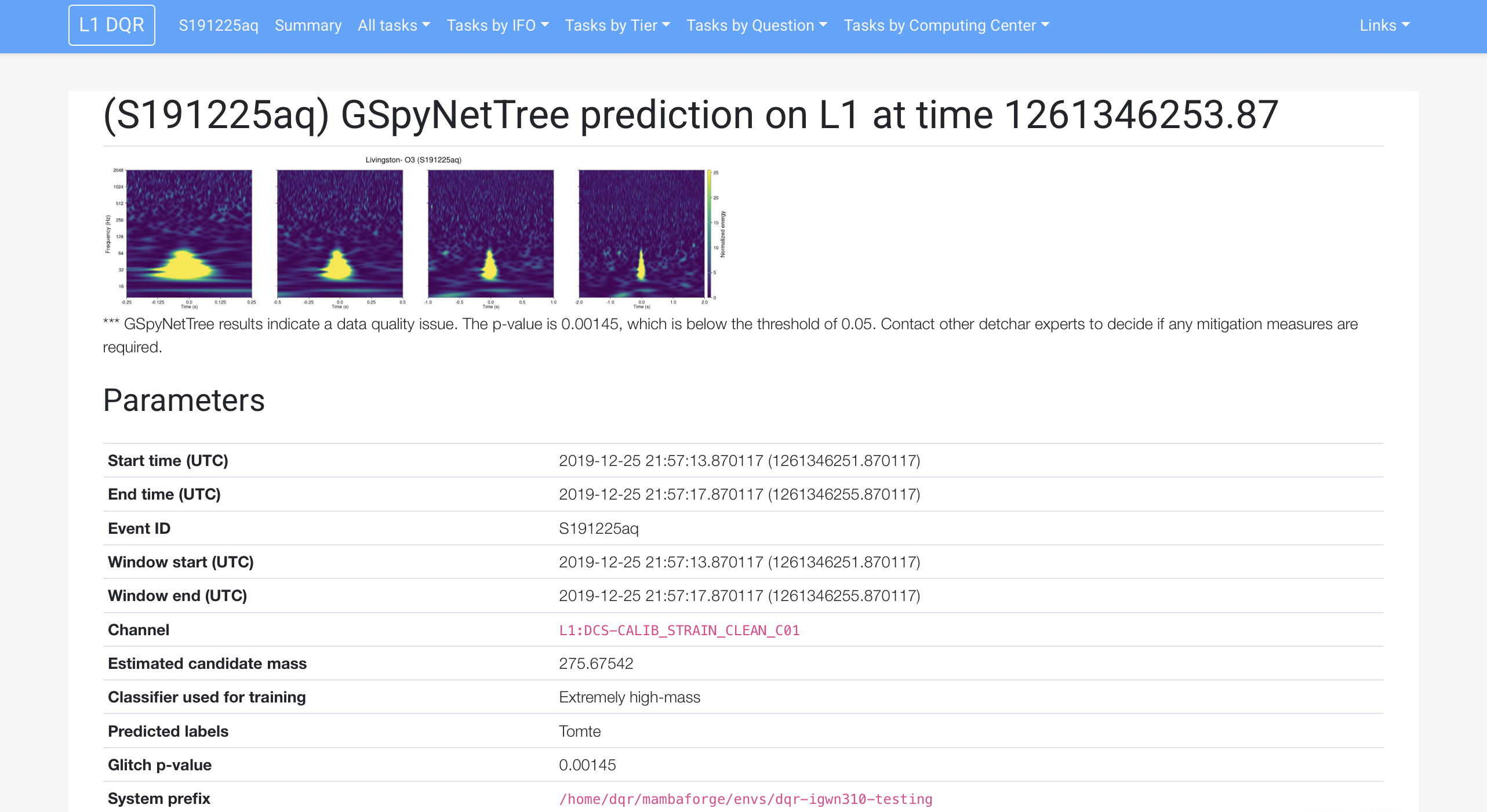}
\caption{An example result page for the GSpyNetTree task.
The displayed candidate is S191225aq, which was retracted due to a similarity between the data from this candidate in LIGO Livingston and a common ``tomte'' glitch~\cite{LIGO_O3_detchar}.}
\label{fig:GSpyNetTree}
\end{figure}

An example of this task run on \red{S191225aq} is shown in Figure~\ref{fig:GSpyNetTree}.
GSpyNetTree is able to flag a morphological similarity between the data from this candidate in LIGO Livingston and a common ``tomte'' glitch~\cite{LIGO_O3_detchar}.
In O3, this candidate was retracted based on the qualitative similarity between the data and this glitch class.
With GSpyNetTree, this correlation is identified automatically based on reproducible, machine-learning classifications of the data.

\subsubsection{GlitchAverage} \label{sec:glitch_average}
This task (implemented in the \texttt{dqr-tasks} package as \texttt{dqr-glitch-average}) evaluates if the local glitch rate (as measured by the Omicron pipeline~\cite{Robinet:2020lbf}) is an outlier over recent data~\cite{doleva2022analysis}.
A high glitch rate, inconsistent with recent data, increases the probability that \ac{DQ} issues are impacting the detection or analysis of a \ac{GW} candidate.  

This task queries the last week of Omicron~\cite{Robinet:2020lbf} triggers to determine if the rate of triggers above a given SNR near the \ac{GW} candidate is exceptional. Details on how these triggers are generated can be found in \emph{Robinet et. al}~\cite{Robinet:2020lbf}. Segments of observing time are broken down into smaller segments and an average glitch rate is calculated on each. The distribution of values of this average compared to the average rate of triggers during the time of the \ac{GW} candidate is presented as the main result of this task. 

In addition to the glitch rate observed in each detector across all frequency bands, this task compares the rate of low-, medium- and high-frequency triggers for the past week of data against the relevant rate at the time of the candidate. The moving average of these triggers as a function of time is plotted along with the rate of low-, medium-, and high-frequency triggers at the candidate time to assist researchers in analyzing the candidate in question.

\subsubsection{Lockcheck} \label{sec:lockcheck}
This task (implemented in the \texttt{dqr-tasks} package as \texttt{dqr-lockcheck}) is designed to verify that all interferometers with data used to analyze a reported event are marked as observing (colloquially referred to as the detector being ``locked''). This task compares the list of interferometers used to identify a candidate and to the list of interferometers in observing mode to identify any use of data from a detector not in observing mode.
When an analysis generates an alert for a potential event, this task compares the set of interferometers listed as detecting the event and against the set of interferometers marked as “observing” at that time. 
If data was used from an interferometer that was not in observing, the task will report a p-value of 0.
In all other cases, the reported p-value is 1.
Therefore, this task can be considered ``binary.''

\subsection{Qualitative tasks}

We note that not all tasks may be suitable for this statistical framework. 
Tasks that can neither report a \textbf{p-value} nor a binary result are considered to be ``qualitative'' tasks.
While it is beneficial to include these qualitative tasks as part of a \texttt{DQRbuild} workflow, the lack of quantification means that these tasks are not to be included in automated checks and are only to be used for events of particular interest.  
However, these qualitative tasks can provide valuable information that is designed to support the interpretation of the statistical tasks.

\subsubsection{Rangecheck} \label{sec:rangecheck}

This task (implemented in the \texttt{dqr-tasks} package as \texttt{dqr-rangecheck}) is designed to report the inspiral range~\cite{Chen:2017wpg} and the sensitivity of the interferometer around the time of a \ac{GW} candidate to check if the detector is stably operating at the time of a candidate.
This task is also useful in cases where one detector may be operational, but is not in observation mode to help evaluate if use of the additional detector data could demonstrably improve the analysis of the candidate in question.

\subsubsection{iDQ Followup} \label{sec:idq_followup}

This task (implemented in the \texttt{dqr-tasks} package as \texttt{dqr-idq-followup}) is a companion task to the main iDQ task described in \ref{sec:idq}. 
If iDQ identifies times of activity in auxiliary channels surrounding a candidate event, the iDQ-followup task visualizes the times identified by iDQ by plotting both spectrograms of the relevant channels alongside ``glitch-grams'' that show the time-frequency location of the SNAX triggers that were used as input to the iDQ algorithm.
These spectrograms and glitch-grams allow manual inspection of the correlated data to validate whether the correlation is spurious or not.

\section{O3 Results}

To validate the performance of the \texttt{DQRbuild} toolkit, we conducted a retrospective analysis of 79 candidates identified as significant in O3 during low-latency data analysis. 
The analyzed list of events includes all candidates that were subject to human vetting in low latency, regardless of whether the candidate was later retracted. 
The inclusion of retracted candidates is desirable for this test, since a key goal of the \texttt{DQRbuild} toolkit is to identify non-astrophysical candidates, such as these. 
If we assume that the data quality in each detector is independent, this would amount to independent 237 tests of the toolkit (1 per interferometer per event). 
Of these 237, we exclude 33 cases where an interferometer was not in observing mode at the time of a candidate. 
This leaves 204 cases considered in this analysis. 

For each candidate and interferometer considered, we ran a suite of tasks similar to those in O4 and compared the set of candidates with identified \ac{DQ} issues against those with \ac{DQ} issues identified during manual vetting. 
Our manual vetting identifies \ac{DQ} issues in 52/204 cases.
Full details of the technical analysis setup are included in \ref{app:setup}.
Since it has previously been shown that O3 validation results included inconsistencies across the dataset regarding what constituted a data quality issue~\cite{Vazsonyi:2022jul}, we manually re-vetted all candidates in the dataset. 
The list of candidates, our manual vetting results for each candidate, and the results from the \texttt{DQRbuild} toolkit for each candidate are shown in \ref{app:DQ-conclusions}.

\subsection{Overall results}

\begin{figure}[t]
\centering
\includegraphics[width=0.85\textwidth]{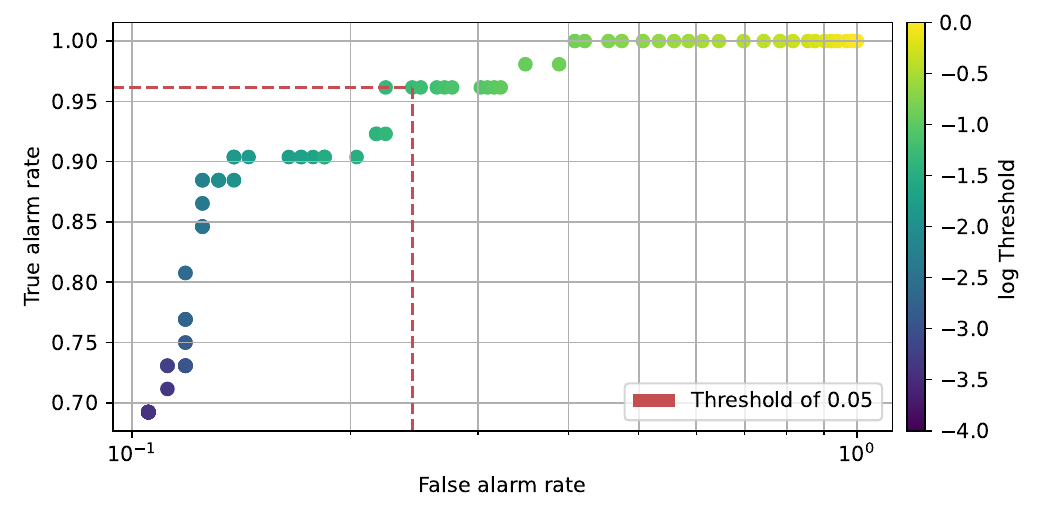}
\caption{The receiver operating characteristic curve for all tasks in the \texttt{DQRbuild} toolkit with respect to different chosen task p-value thresholds.
The color of each point indicates the logarithm of the p-value threshold used to identify alarms.  
The false alarm rate and true alarm rate for the threshold used in most of this work (0.05) is indicated with red dashed lines. 
Note that the chosen p-value threshold is not equal to the false alarm rate measured using O3 candidates.}
\label{fig:roc}
\end{figure}

Overall, we find that, assuming a p-value threshold of \red{0.05} for each task, the \texttt{DQRbuild} toolkit identifies a \ac{DQ} issue in \red{50}/\red{52} cases (on a per-interferometer basis), while falsely identifying \ac{DQ} issues in \red{37}/\red{152} cases. 
This corresponds to a false alarm rate of \red{24\%} and a true alarm rate of \red{96\%}. 
This threshold is user-configurable and significantly affects the true and false alarm rates. 
How these metrics change as a function of the threshold is shown in Figure~\ref{fig:roc}.
The chosen threshold of 0.05 is marked with red dashed lines.
We find that the true alarm rate greatly exceeds the false alarm rate for any chosen threshold, demonstrating that the \texttt{DQRbuild} toolkit is effective at identifying relevant \ac{DQ} issues. 
However, the true alarm rate drops below \red{75\%} when the false alarm rate reaches \red{10\%}.
This relatively high false alarm rate can be explained by tasks flagging non-Gaussian features that we do not mark as \ac{DQ} issues.
Differences between the datasets used to tune the task p-values and the chosen O3 dataset may also contribute.

\begin{figure}[t]
\centering
\includegraphics[width=0.45\textwidth]{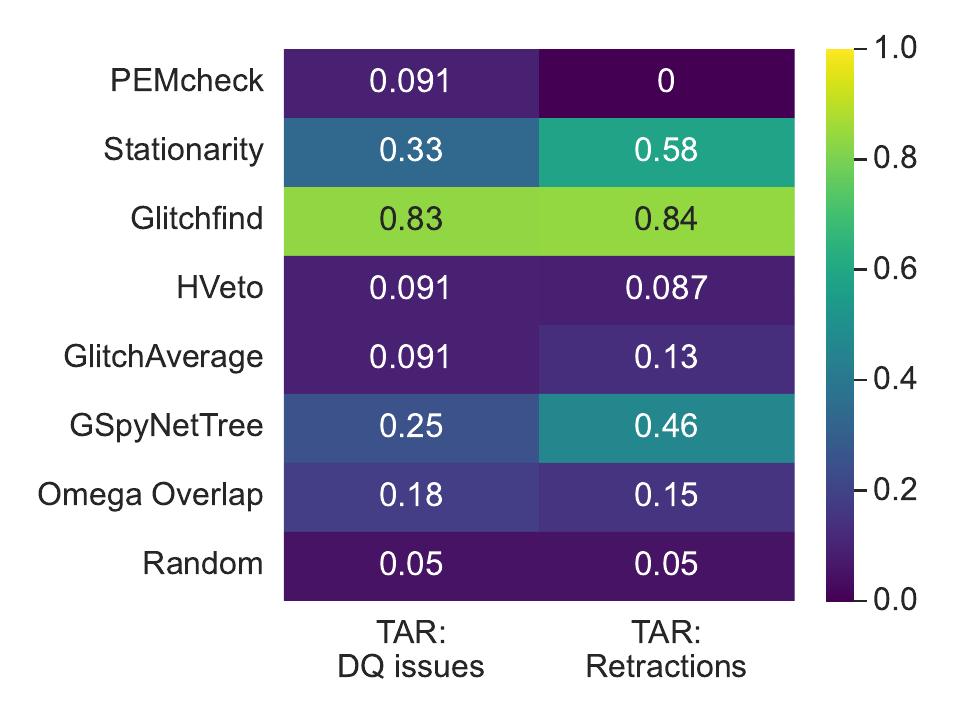}
\includegraphics[width=0.45\textwidth]{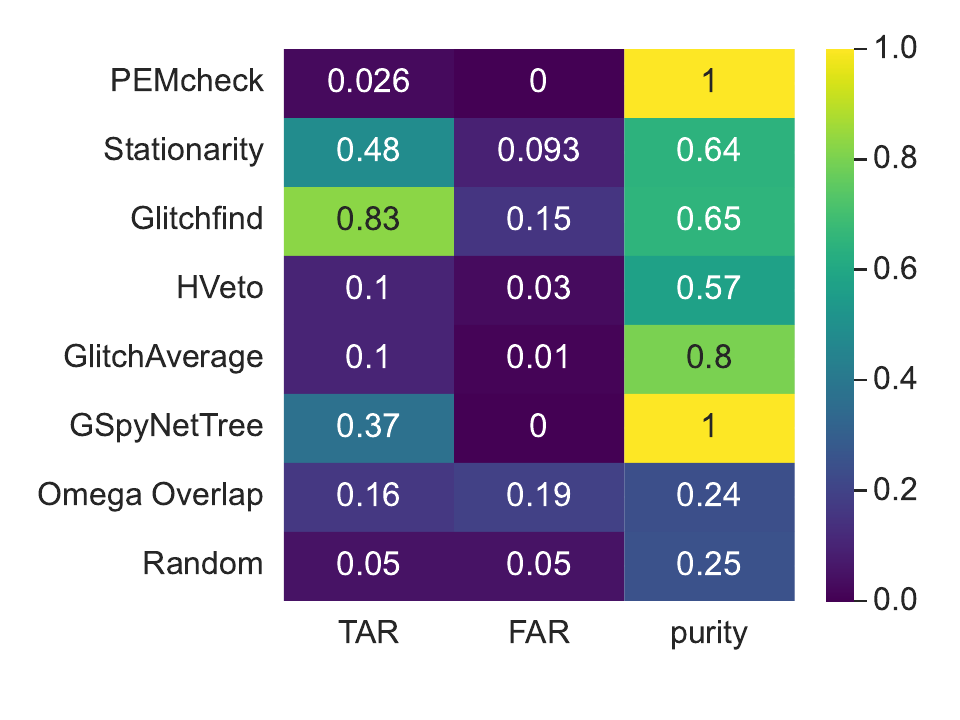}
\caption{Results from O3 for each individual statistical task considered in this analysis. Two separate heatmaps are shown. The left panel shows the true alarm rate (TAR) for either candidates labeled as having \ac{DQ} issues or being retracted as part of the O3 low-latency vetting. The right shows the true alarm rate (TAR), the false alarm rate (FAR), and purity at flagging \ac{DQ} issues for all tasks in this analysis. The panels also show statistics expected of a random number generator. Note that each panel uses a different color bar.}
\label{fig:stats}
\end{figure}

Looking at the performance of individual tasks, we find significant variability in the accuracy and precision across the tasks in the \texttt{DQRbuild} toolkit.
A number of per-task metrics are displayed in Figure~\ref{fig:stats}. 
The first is the true alarm rate, false alarm rate, and fraction of all alarms that are true alarms (the purity of alarms) for each task over the entire analysis. 
The second is the accuracy for cases with no \ac{DQ} issues, \ac{DQ} issues, and retractions. 
For comparison, we include the results that would be obtained if a task were randomly assigned p-values uniformly between 0 and 1.

We find that a small number of tasks dominate the flagged \ac{DQ} issues, while others reported few, if any, \ac{DQ} issues.
Almost all flagged \ac{DQ} issues were done by the Stationarity, Glitchfind, GSpyNetTree, and omega overlap tasks. 
Other tasks, such as PEMCheck, HVeto, and GlitchAverage, have high purity, but low accuracy in finding \ac{DQ} issues.
We further investigate the performance of these two groups separately.

\subsection{Follow-up investigations of tasks with high true positive rates}

Among the tasks that flagged more than \red{10\%} of \ac{DQ} issues, the false alarm and true alarm rate of these tasks greatly varied.
Assuming that the chosen threshold corresponds to a false alarm probability of \red{5\%}, the Stationarity, Glitchfind, and Omega Overlap tasks report \red{2-4} times higher false alarm rate than expected.
Conversely, the GSpyNetTree task reported \red{no} false alarms for the dataset and the specified threshold. 
For true alarms, Glitchfind identified \red{83\%} of all \ac{DQ} issues, while the Stationarity, GSpyNetTree, and omega overlap tasks identified \red{48\%}, \red{37\%}, and \red{16\%}, respectively.
The Glitchfind and GSpyNetTree tasks are thus the two that outperform all others in terms of FAR and TAR across all tasks that identify a non-trivial number of \ac{DQ} issues. 
We find that Glitchfind is most likely to identify a real \ac{DQ} issue, but GSpyNetTree has the highest purity. 

Conversely, we find that the false alarm rate exceeds the true alarm rate in the omega overlap task. 
We therefore exclude the omega overlap task from the remaining analyses, as this analysis suggests that this task (as configured) is not able to differentiate cases with and without \ac{DQ} issues.

\subsection{Follow-up investigations of tasks with low true positive rates}

A significant number of tasks rarely flagged any data quality issues, but generally had high purities when they did. These tasks include PEMcheck, GlitchAverage, and HVeto. 
If we examine the specific goals of each task, we can understand that this behavior is expected. 

The tasks PEMcheck and HVeto are designed to identify correlations between the strain channel and auxiliary channels. 
Neither statistic is a p-value based on Gaussian noise; instead, it is used to estimate the significance of any identified correlations. 
If no such correlations exist, then the rate of alarms from these tasks is expected to be low. 
In O3, only \red{3} candidates were retracted or flagged based on auxiliary channel information, and \red{all} of these cases were found by one of these two tasks. 
We conclude that both tasks are performing as desired and can flag these types of issues appropriately. 

The final task, GlitchAverage, identifies cases where the glitch rate is abnormally high.
Since this p-value is empirically determined from real LIGO data, it is expected that the task will flag data quality issues only 5\% of the time (assuming a p-value threshold of 0.05). 
This rate \red{is consistent} with what is observed in our analysis. 

These tasks highlight that the overall rate of true alarms from a task is not the only determining factor when choosing the set of tasks to run in the data quality report. 
While care must be taken not to include duplicate tasks (which would only increase the rate of false alarms), tasks designed to address specific, rare data quality issues can be important to include in the validation process to ensure these issues are not present.

\subsection{Retrospective full automation test}

The eventual goal of the \texttt{DQRbuild} toolkit is not just to automate the analysis of data quality issues surrounding \ac{GW} events, but also to automate the decision-making process related to these issues. 
As an example of how this could be structured, we conduct a retrospective test of what complete automation of the event validation process would have looked like if the data quality report toolkit had been used in O3. 
Although there are three distinct actions taken in low-latency analyses (retraction of a candidate, flagging data quality issues around a candidate, and no issues with a candidate), the \texttt{DQRbuild} toolkit is designed only to differentiate cases with or without data quality issues. 
We therefore also investigate the results if only part of the workflow was automated, allowing humans to differentiate between different types of data quality issues. 

\begin{figure}[t]
\centering
\includegraphics[width=0.9\textwidth]{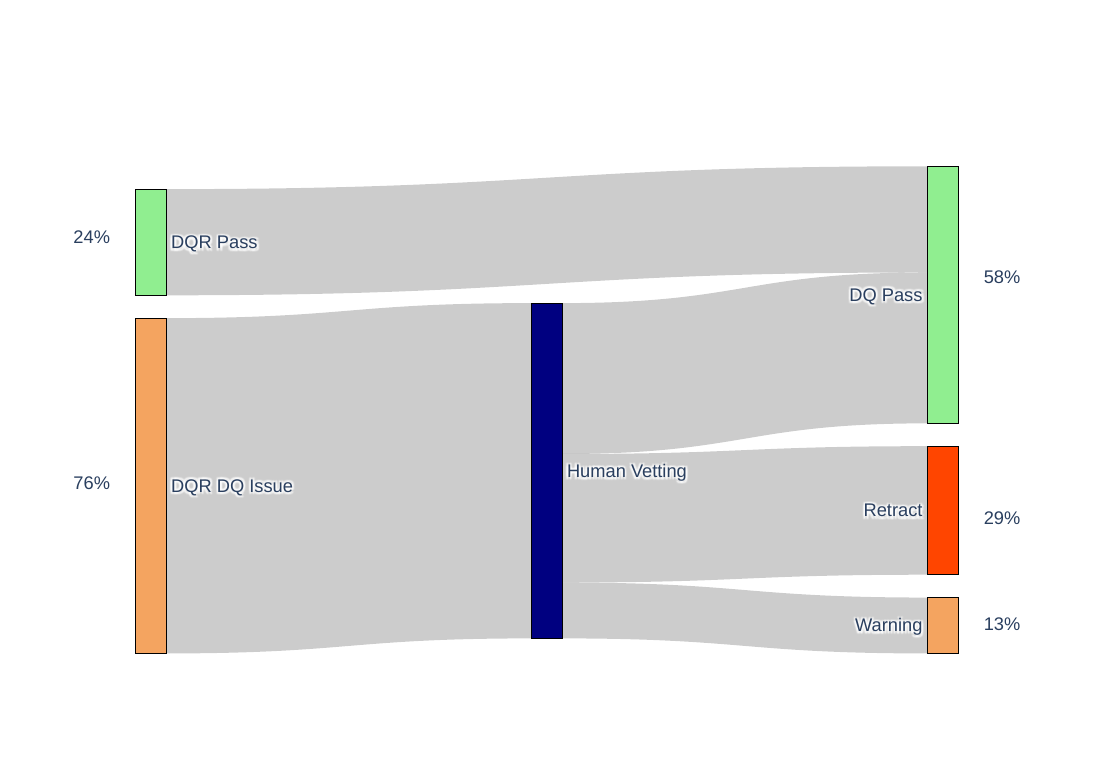}
\caption{Sankey diagram~\cite{otto2022overview} of how the \texttt{DQRbuild} toolkit could have been used in a semi-autonomous fashion to reduce the amount of human vetting required to process low latency candidates. In this diagram, all candidates with \ac{DQ} issues identified by the \texttt{DQRbuild} toolkit are passed to an additional human vetting stage. 
The candidate is then manually sorted into the ``DQ Pass,'' ``Retract,'' or ``DQ Warning'' categories. 
Candidates without any identified \ac{DQ} issues by the \texttt{DQRbuild} toolkit are automatically assigned the ``DQ Pass'' label. The displayed percentages indicate the fraction of the total dataset assigned the relevant label in the Sankey diagram. }
\label{fig:sankey}
\end{figure}

The results of this investigation can be seen in Figure~\ref{fig:sankey}.
In the case that \texttt{DQRbuild} toolkit results are used to identify \ac{DQ} issues autonomously, we find that \red{100\%} of \ac{DQ} issues (both \ac{DQ} warnings and retractions) are successfully flagged. 
We also find that cases with no issues are classified correctly \red{41\%} of the time.
Although the \texttt{DQRbuild} toolkit can flag every \ac{DQ} issue, the rate of false alarms means these results do not align with the philosophy of not retracting or flagging issues with candidates incorrectly. 
We conclude that use of the \texttt{DQRbuild} toolkit in this way would require a change in how users interpret results from the LVK. 

To address this high rate of false alarms, we investigate the impact of adding a level of human inspection to ``catch'' these misclassifications. 
This manual vetting would also allow separating cases where a retraction is needed from those where only a \ac{DQ} warning should be issued. 
We demonstrate what this would look like in practice in Figure~\ref{fig:sankey}.
We can see that \red{45\%} of cases where the \texttt{DQRbuild} toolkit flagged a \ac{DQ} issue, the human would have overridden the result. 
Conversely, \red{100\%} of all \ac{DQ} issues were correctly identified; the impact of human vetting would be to correctly sort these cases between retraction and \ac{DQ} warning rather than overriding \texttt{DQRbuild} toolkit conclusions. 
In total, \red{24\%} of candidates would require no human input in this scenario, significantly reducing the human cost of follow-up for low-latency searches of \ac{GW} data.

\section{Discussion}

The primary goal of the \texttt{DQRbuild} toolkit is to automate the \ac{DQ} vetting of \ac{GW} candidates with high accuracy. 
Given the \red{96\%} true alarm rate for tasks, we have shown that this toolkit can quickly and accurately identify the vast majority of \ac{DQ} issues that can impact low-latency data analysis in ground-based \ac{GW} interferometers. 
However, as we move towards the next observing run, there are several limitations in the current version of the \texttt{DQRbuild} toolkit that should be addressed.

Significant research and development must be undertaken to move from the current focus of the \texttt{DQRbuild} toolkit on the automation of analyses to the automation of decisions. 
Although the toolkit is able to effectively identify \ac{DQ} issues, it does not guide users on what is needed to address them. 
Current tools rely on an additional stage of human vetting, whether for all candidates or for candidates with identified \ac{DQ} issues, before any final conclusion is made. 
Automation of these processes should ideally also account for the potential for novel astrophysical scenarios, such as overlapping events~\cite{Janquart:2022fzz}, \acp{GW} not consistent with general relativity~\cite{Gupta:2024gun}, or unexpected signal morphologies.

Another limitation of the current toolkit is the relatively high false alarm rate. 
This can partially be attributed to the highly conservative approach used when designing these tools; 
a true alarm rate close to unity should be considered a requirement.
This approach is currently acceptable, since there has always been a human vetting stage in the low-latency analysis process that can manually override any flagged \ac{DQ} issues. 
If there is a desire to move towards full automation, a high rate of false alarms from automated analysis would no longer be acceptable.
There should therefore be a decision on how to balance the human cost of vetting against the desired purity of the automated results.

In this new automation paradigm, it is also important that we balance the costs and benefits of the number of tasks that are used in the \texttt{DQRbuild} toolkit. 
Although more science tasks will increase the rate at which \ac{DQ} issues are identified, 
more tasks also results in a higher false alarm rate, limiting the benefits of these additional tasks.
Complicating this calculation is the wide range of \ac{DQ} issues that tasks are designed to target.
Some tasks may be designed to only flag \ac{DQ} issues that are rarely present;
a low true alarm rate for real candidates is not a sufficient reason not to include the task in the \texttt{DQRbuild} toolkit.
A deep understanding of the goals of each task is needed to optimally balance the rate of false alarms and the overall coverage of the \texttt{DQRbuild} toolkit when considering adding or removing an individual task.

The \texttt{DQRbuild} toolkit exposes one of the significant challenges when conducting a global \ac{GW} analysis: data access. 
Hundreds of individual data streams, along with many other processed datasets from each interferometer in the global network, are required for low-latency \ac{DQ} vetting of \ac{GW} candidates. 
Accessing this data, performing the required analysis, and then reaching a centralized conclusion about a candidate are among the most significant stressors on the currently developed \ac{GW} data distribution system. 
Future improvements in the robustness and speed of vetting will depend on the hardening of these data distribution systems across the global network.

Communication of results presents an additional challenge for the effectiveness of low-latency vetting beyond the technical challenges already described. 
Similar to the technical challenges, the desire for more speed and increased automation will also stress this aspect of the scientific endeavor.
Effective \ac{DQ} vetting of candidates requires a thorough understanding of the instrumentation in every \ac{GW} interferometer in the network. 
However, the primary consumers of these results are other astronomers with expertise in fields other than \ac{GW} detectors.
At present, the translation of results between experts and non-experts is done manually for each candidate.
If humans are removed from the loop, carefully planning out this communication in advance will be an important part of that process.

In addition to the testing on O3 data, this toolkit has been used (in conjunction with the \texttt{VirgoDQR}~\cite{Virgo_O3_detchar_tools}, \texttt{GWDetchar}~\cite{gwdetchar}, and \texttt{event-validation}~\cite{LIGO:2024kkz} packages) to validate all \ac{GW} candidates identified in LIGO-Virgo-KAGRA data by low-latency detection pipelines during O4. 
Validation results for announced candidates based on this toolkit have also been included alongside other previously announced astrophysical results~\cite{GW231123,GW250114,GWTC-4}. 
Once results from all O4 candidates are available, these will be an important source of data for future improvements to the \texttt{DQRbuild} toolkit.
Additional details on the analysis configuration of the \texttt{DQRbuild} toolkit used in O4 and a holistic look at the performance of the \texttt{DQRbuild} toolkit based on \ac{GW} events from this observing run are expected in future publications.

All packages that comprise the \texttt{DQRbuild} toolkit and are discussed in this work are open source and available on \url{https://git.ligo.org}~\cite{dqrbuild-gitlab}.

\section{Acknowledgments}

We thank the LIGO-Virgo-KAGRA detector characterization groups for helpful discussions and input as well as Laura Nuttall for comments during internal review of this manuscript. 
This work was supported by a grant from the Simons Foundation International [SFI-MPS-SSRFA-00023625, DD].
ZY is supported by NSF award PHY-2409740.
JA is supported by NSF award PHY-2110594.
AHC is supported by NSF award PHY-2409502.

This material is based upon work supported by NSF's LIGO Laboratory which is a major facility fully funded by the National Science Foundation, as well as the Science and Technology Facilities Council (STFC) of the United Kingdom, the Max-Planck-Society (MPS), and the State of Niedersachsen/Germany for support of the construction of Advanced LIGO and construction and operation of the GEO600 detector. Additional support for Advanced LIGO was provided by the Australian Research Council. Virgo is funded, through the European Gravitational Observatory (EGO), by the French Centre National de Recherche Scientifique (CNRS), the Italian Istituto Nazionale di Fisica Nucleare (INFN) and the Dutch Nikhef, with contributions by institutions from Belgium, Germany, Greece, Hungary, Ireland, Japan, Monaco, Poland, Portugal, Spain. KAGRA is supported by Ministry of Education, Culture, Sports, Science and Technology (MEXT), Japan Society for the Promotion of Science (JSPS) in Japan; National Research Foundation (NRF) and Ministry of Science and ICT (MSIT) in Korea; Academia Sinica (AS) and National Science and Technology Council (NSTC) in Taiwan.

LIGO was constructed by the California Institute of Technology 
and Massachusetts Institute of Technology with funding from 
the National Science Foundation, 
and operates under cooperative agreement PHY-1764464. 
Advanced LIGO was built under award PHY-0823459.
The authors are grateful for computational resources provided by the 
LIGO Laboratory and supported by 
National Science Foundation Grants PHY-0757058 and PHY-0823459.
This work carries LIGO document number P2600016.

\paragraph{List of Acronyms}

\begin{acronym}[DQR]
  \acro{DQ}{Data quality}
  \acro{DQR}{Data quality report}
  \acro{GW}{Gravitational wave}
\end{acronym}

\appendix

\section{Science task set up} \label{app:setup}

As part of our testing on O3 data, we ran the following tasks:

\begin{itemize}
    \item HVeto (Section~\ref{sec:hveto})
    \item Stationarity (Section~\ref{sec:stationarity})
    \item PEMcheck (Section~\ref{sec:pemcheck})
    \item Omega Overlap (Section~\ref{sec:omega_overlap})
    \item Glitchfind (Section~\ref{sec:glitchfind})
    \item GSpyNetTree (Section~\ref{sec:gspynettree})
    \item GlitchAverage (Section~\ref{sec:glitch_average})
    \item Lockcheck (Section~\ref{sec:lockcheck})
    \item Rangecheck (Section~\ref{sec:rangecheck})
    \item GWDetChar-Omegascan (from \texttt{GWDetChar} package~\cite{gwdetchar})
\end{itemize}

Note that some of these tasks are not mentioned in the discussion of the results in the main body of the text.
The Lockcheck task is binary, and is designed to identify candidates that were found out of observing. 
No candidates from O3 fit the description, so this task was excluded from discussions.
The Rangecheck and GWDetChar-Omegascan tasks are qualitative tasks and, hence, return no statistical information. 
For this reason, these tasks are also excluded from discussions in the main text. 

The iDQ and iDQ followup tasks that are described in the main text, but are not included in our analysis of O3 events as the required iDQ data was not represented in the same way during O3 and it is in O4. 
We therefore are not able to test the O4-version of the iDQ task on O3 events. 

Due to limitations on what data was available at the computing clusters where this analysis was conducted, not all tasks were run on all detectors. 
Tasks that utilize only strain and state vector data (GlitchAverage, GSpyNetTree, Glitchfind, Stationarity, Lockcheck, Rangecheck, and GWDetChar-Omegascan) were tested with LIGO Hanford, LIGO Livingston, and Virgo data. 
The Omega Overlap, HVeto and PEMCheck tasks were only tested with LIGO Hanford and LIGO Livingston data due to differences in data availability between the two LIGO detectors and the Virgo detector.
These strain-based tasks have been previously tested on KAGRA data, but data from this interferometer is not included in this analysis as no KAGRA data was available during times of identified \ac{GW} candidates in O3.

\section{O3 DQ conclusions} \label{app:DQ-conclusions}

As part of this analysis, we manually re-analyzed the \ac{DQ} on a per-interferometer basis for all significant low-latency candidates identified in O3.
This process was completed to ensure that the standard for manually identifying \ac{DQ} issues was consistent across the entire dataset. 
We use these updated results when comparing the accuracy of an automated \texttt{DQRbuild} workflow against manual results rather than the results reported by the LIGO-Virgo-KAGRA collaboration in notices released publicly during O3, since previous investigations have found significant differences in how different individuals manually validate \ac{GW} candidates~\cite{Vazsonyi:2022jul}.
As manual vetting has an element of subjectivity, we expect some variation between our categorizations listed in Tables~\ref{tab:o3results1} and~\ref{tab:o3results2} and the original O3 results or the same procedure conducted by other experts.  
Our conclusions, along with the statistical results of the \texttt{DQRbuild} toolkit and the result based on notices sent as part of the O3 low-latency campaign are explicitly shown in Tables~\ref{tab:o3results1} and~\ref{tab:o3results2}.

\begin{table}[htbp]
\centering
\begin{tabular}{l|c|ccc}
\hline
Candidate & O3 conclusion & H1 result & L1 result & V1 result \\
\hline
S190408an & - & - & - & - \\
S190412m & - & - & - & - \\
S190421ar & - & - & - & \XSolid \\
S190426c & \ac{DQ} issue & - & \ac{DQ} issue & - \\
S190503bf & \ac{DQ} issue & - & \ac{DQ} issue & - \\
S190510g & \ac{DQ} issue & - & \ac{DQ} issue & - \\
S190512at & - & - & - & - \\
S190513bm & - & - & - & - \\
S190517h & - & - & - & - \\
S190518bb & Retract & \ac{DQ} issue & - & - \\
S190519bj & - & - & - & - \\
S190521g & - & - & - & - \\
S190521r & - & - & - & \XSolid \\
S190524q & Retract & - & \ac{DQ} issue & - \\
S190602aq & - & - & - & - \\
S190630ag & - & - & - & \ac{DQ} issue \\
S190701ah & \ac{DQ} issue & - & \ac{DQ} issue & - \\
S190706ai & - & - & - & - \\
S190707q & - & - & - & \XSolid \\
S190718y & - & \ac{DQ} issue & \ac{DQ} issue & - \\
S190720a & - & - & - & - \\
S190727h & - & - & - & - \\
S190728q & - & - & - & - \\
S190808ae & Retract & - & \ac{DQ} issue & \ac{DQ} issue \\
S190814bv & \ac{DQ} issue & \XSolid & \ac{DQ} issue & - \\
S190816i & Retract & \XSolid & \ac{DQ} issue & - \\
S190822c & Retract & \XSolid & \ac{DQ} issue & \ac{DQ} issue \\
S190828j & - & - & - & - \\
S190828l & - & - & - & - \\
S190829u & Retract & \XSolid & \ac{DQ} issue & - \\
S190901ap & - & \XSolid & - & - \\
S190910d & - & - & - & - \\
S190910h & - & \XSolid & - & \XSolid \\
S190915ak & - & - & - & - \\
S190923y & - & - & \ac{DQ} issue & \XSolid \\
S190924h & - & - & - & - \\
S190928c & Retract & - & \ac{DQ} issue & - \\
S190930s & - & - & - & \XSolid \\
S190930t & - & \XSolid & \ac{DQ} issue & \XSolid \\
\hline
\end{tabular}
\caption{Table showing the event validation results from low latency vetting of candidates in O3 and the re-vetted results used in this work. Note that the O3 conclusion is given as a single result, while the re-vetted results are provided for each interferometer. Cases where no \ac{DQ} issue was identified are marked with ``-'', while cases where an interferometer was offline are marked with ``\XSolid''. Additional results are displayed in Table~\ref{tab:o3results2}. }
\label{tab:o3results1}
\end{table}

\begin{table}[htbp]
\centering
\begin{tabular}{l|c|ccc}
\hline
Candidate & O3 conclusion & H1 result & L1 result & V1 result \\
\hline
S191105e & \ac{DQ} issue & - & \ac{DQ} issue & - \\
S191109d & \ac{DQ} issue & \ac{DQ} issue & - & \XSolid \\
S191110af & Retract & \ac{DQ} issue & \ac{DQ} issue & - \\
S191110x & Retract & - & \ac{DQ} issue & - \\
S191117j & Retract & - & \ac{DQ} issue & \XSolid \\
S191120aj & Retract & - & \ac{DQ} issue & \XSolid \\
S191120at & Retract & - & \ac{DQ} issue & \ac{DQ} issue \\
S191124be & Retract & \XSolid & - & \ac{DQ} issue \\
S191129u & - & - & - & \XSolid \\
S191204r & - & - & - & \XSolid \\
S191205ah & - & - & - & - \\
S191212q & Retract & \ac{DQ} issue & - & \XSolid \\
S191213ai & Retract & - & \ac{DQ} issue & \ac{DQ} issue \\
S191213g & \ac{DQ} issue & \ac{DQ} issue & \ac{DQ} issue & - \\
S191215w & - & - & - & - \\
S191216ap & - & - & \XSolid & - \\
S191220af & Retract & \XSolid & - & \ac{DQ} issue \\
S191222n & - & - & - & - \\
S191223ao & - & - & - & - \\
S191225aq & Retract & \XSolid & \ac{DQ} issue & \ac{DQ} issue \\
S200105ae & - & \XSolid & - & - \\
S200106au & Retract & \ac{DQ} issue & \ac{DQ} issue & - \\
S200106av & Retract & \ac{DQ} issue & \ac{DQ} issue & - \\
S200108v & Retract & - & \ac{DQ} issue & - \\
S200112r & - & \XSolid & - & - \\
S200114f & - & - & - & - \\
S200115j & \ac{DQ} issue & - & \ac{DQ} issue & - \\
S200116ah & Retract & - & \ac{DQ} issue & \XSolid \\
S200128d & - & - & - & \XSolid \\
S200129m & \ac{DQ} issue & - & \ac{DQ} issue & \ac{DQ} issue \\
S200208q & - & - & - & - \\
S200213t & - & - & - & \ac{DQ} issue \\
S200219ac & - & - & \ac{DQ} issue & - \\
S200224ca & - & - & \ac{DQ} issue & - \\
S200225q & - & - & - & \XSolid \\
S200302c & - & - & \XSolid & - \\
S200303ba & Retract & - & - & \ac{DQ} issue \\
S200308e & Retract & - & \ac{DQ} issue & - \\
S200311bg & - & - & - & - \\
S200316bj & - & - & - & \ac{DQ} issue \\
\hline
\end{tabular}
\caption{Extension of Table~\ref{tab:o3results1}, showing the event validation results from low latency vetting of candidates in O3 and the re-vetted results used in this work. Note that the O3 conclusion is given as a single result, while the re-vetted results are provided for each interferometer. Cases where no \ac{DQ} issue was identified are marked with ``-'', while cases where an interferometer was offline are marked with ``\XSolid''.}
\label{tab:o3results2}
\end{table}

Our \ac{DQ} results for each interferometer are split into either an identified \ac{DQ} issue or no \ac{DQ} issue (labeled ``-'').
The O3 notices split conclusions into either retractions (labeled ``Retract''), \ac{DQ} warnings (labeled ``DQ issue''), or no \ac{DQ} issues (labeled ``-'').
Hence, a retraction or \ac{DQ} warning can be considered consistent with what we label a \ac{DQ} issue.

\bibliographystyle{my-iopart-num}
\bibliography{references}

\end{document}